\documentclass{mem}
\usepackage{natbib}\usepackage{txfonts}\usepackage{balance}
\usepackage{graphicx}
\usepackage[a4paper]{hyperref}
\idline{75}{1}
\begin{document}


\def\eq#1{\begin{equation} #1 \end{equation}}
\def\E#1{\hbox{$10^{#1}$}}
\def\about {\hbox{$\sim$}}
\def\la    {\hbox{$\lesssim$}}
\def\ga    {\hbox{$\gtrsim$}}
\def\x     {\hbox{$\times$}}
\def\half  {\hbox{$\frac12$}}
\def\mic   {\hbox{$\mu$m}}
\def\deg   {\hbox{$^\circ$}}
\def\Lo     {\hbox{$L_{\odot}$}}
\def\LEdd   {\hbox{$L_{\rm Edd}$}}
\def\Mo     {\hbox{$M_{\odot}$}}
\def\tV    {\hbox{$\tau_{\rm V}$}}
\def\nc    {\hbox{$n_{\rm cl}$}}
\def\Rd     {\hbox{$R_{\rm d}$}}
\def\Ro     {\hbox{$R_{\rm o}$}}
\def\Rc     {\hbox{$R_{\rm cl}$}}
\def\Rx     {\hbox{$R_{\rm x}$}}
\def\R     {\hbox{$\cal R$}}
\def\erg   {\hbox{erg\,s$^{-1}$}}
\def\kms   {\hbox{km\,s$^{-1}$}}
\def\MBH   {\hbox{$M_{\bullet\,7}$}}
\def\Mo     {\hbox{$M_{\odot}$}}
\def\Mc    {\hbox{$M_{\rm cl}$}}
\def\NH     {\hbox{$N_{\rm H}$}}
\def\cc    {\hbox{cm$^{-3}$}}
\def\cs    {\hbox{cm$^{-2}$}}


\title{Disk-outflow Connection and the \\
       Molecular Dusty Torus}

\author{Moshe Elitzur}

\institute{ Physics \& Astronomy Dept., University of Kentucky, Lexington, KY
40506-0055, USA \email{moshe@pa.uky.edu}
}

\authorrunning{Elitzur}

\titlerunning{The AGN Torus}

\abstract{Toroidal obscuration is a keystone of AGN unification. There is now
direct evidence for the torus emission in infrared, and possibly water masers.
Here I summarize the torus properties, its possible relation to the immediate
molecular environment of the AGN and present some speculations on how it might
evolve with the AGN luminosity.


}
\maketitle{}

\section{Introduction}

Although there are numerous classes of active galactic nuclei (AGN), a unified
scheme has been emerging steadily \citep[e.g.,][]{Ski93, Urry95}. The nuclear
activity is powered by a supermassive (\about\E6--\E{10} \Mo) black hole and
its accretion disk. This central engine is surrounded by a dusty torus, which
can be considered an acronym for Toroidal Obscuration Required by Unification
Schemes: much of the observed diversity is simply explained as the result of
viewing this axisymmetric geometry from different angles. Because of the
anisotropic obscuration of the central region, sources viewed face-on are
recognized as ``type 1'', those observed edge-on are ``type 2''. From basic
considerations, \cite{Krolik88} concluded that the obscuration is likely to
consist of a large number of individually very optically thick dusty clouds.

\section{Torus Properties}

\subsection{Obscuration}
\label{sec:f2}

There are clear indications that the optical depth at visual in the torus
equatorial plane is at least \tV\ \ga\ 10.  If the dust abundance in the torus
is similar to Galactic interstellar regions, the equatorial column density is
at least \NH\ \ga\ 2\x\E{22}~\cs.

\begin{figure}[t]
 \centering\leavevmode
 \includegraphics[width=0.45\hsize,clip]{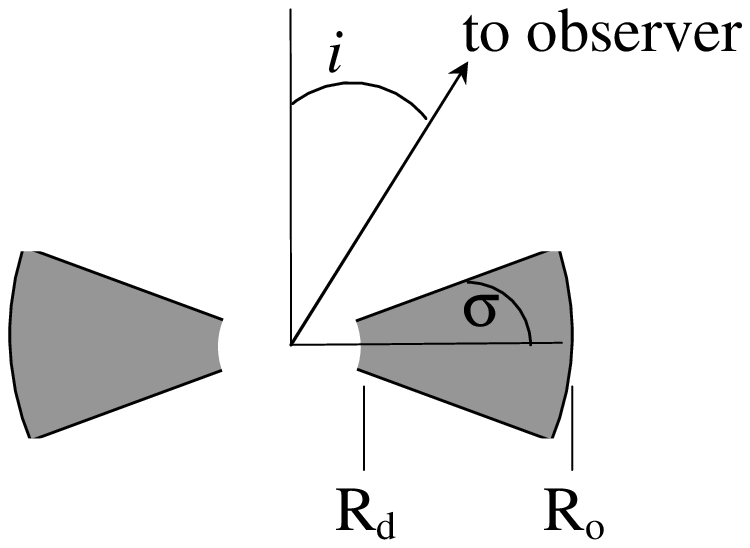} \hfill
 \includegraphics[width=0.45\hsize,clip]{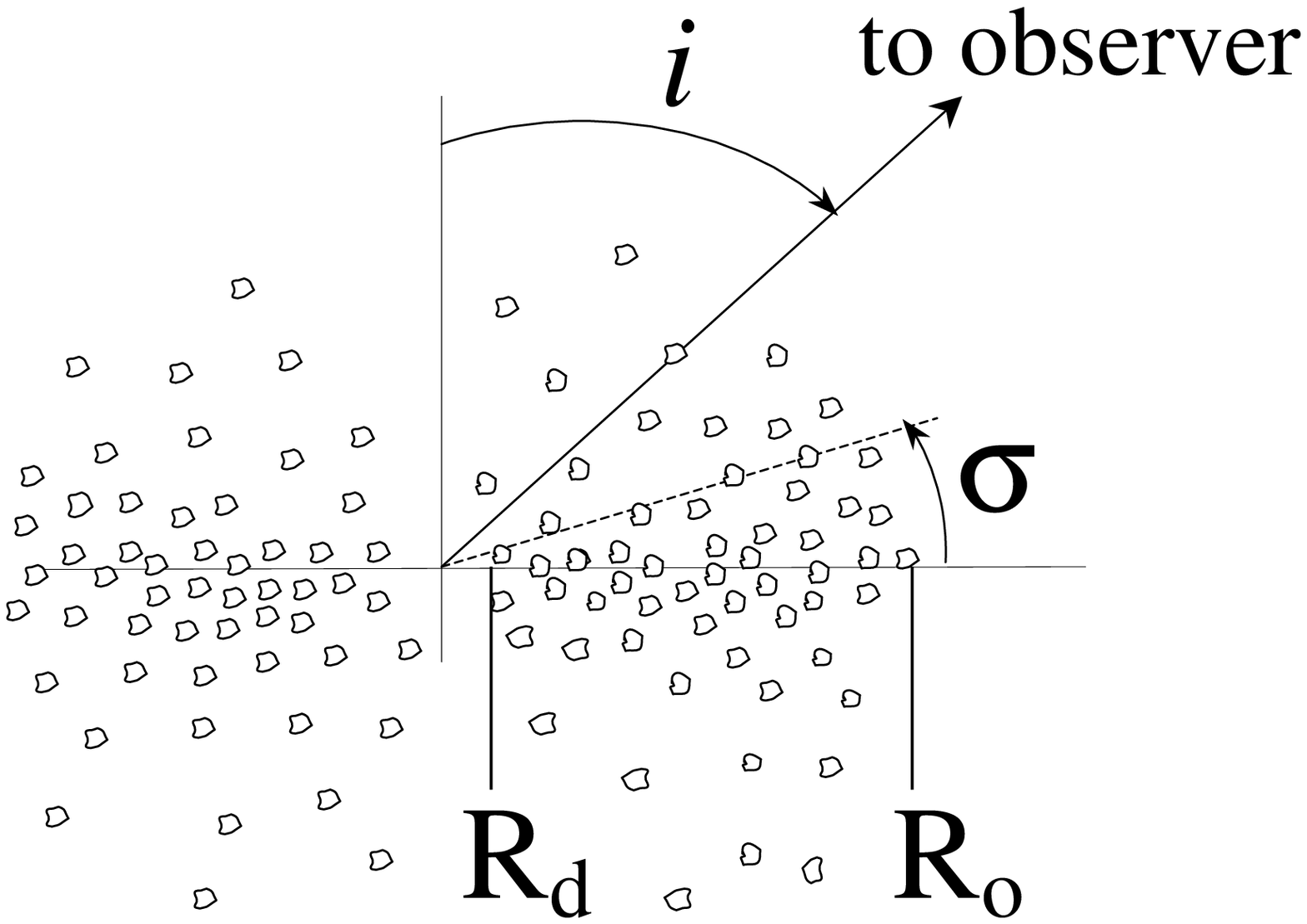}

\caption{AGN classification according to unified schemes. {\em Left}: In a
smooth-density torus, the viewing angle (from the axis) $i = \frac12\pi -
\sigma$ separates between type 1 and type 2 viewing. {\em Right:} In a clumpy,
soft-edge torus, the probability for direct viewing of the AGN decreases away
from the axis, but is always finite.} \label{Fig:Smooth_Clumpy}
\end{figure}

The classification of AGN into types 1 and 2 is based on the extent to which
the nuclear region is visible, therefore source statistics can determine the
angular extent of the torus obscuration. In its standard formulation, the
unification approach posits the viewing angle as the sole factor in determining
the AGN type. This is indeed the case for a smooth-density torus that is
optically thick within the angular width $\sigma$ (figure
\ref{Fig:Smooth_Clumpy}, left sketch). If $f_2$ denotes the fraction of type 2
sources in the total population, then $f_2 = \sin\sigma$. From statistics of
Seyfert galaxies \cite{Schmitt01} find that $f_2 \simeq$ 70\%. and deduce
$\sigma \simeq$ 45\deg. If $H$ denotes the torus height at its outer radius
\Ro, this implies $H/\Ro$ \about\ 1. Taking account of the torus clumpiness
modifies this relation fundamentally, as is evident from the right sketch in
figure \ref{Fig:Smooth_Clumpy}: the obscuration of the central engine becomes a
viewing-angle dependent probability that depends on both the width of the cloud
angular distribution and the number of clouds along the line of sight. In a
Gaussian angular distribution with 5 clouds along radial rays in the equatorial
plane, $f_2$ = 70\% implies $\sigma$ = 33\deg\ and $H/\Ro$ \about\ 0.7
\citep{AGN2}. The clumpy nature of the obscuration also implies that AGN can
occasionally flip between types 1 and 2, as observed \citep[see][and references
therein]{Aret99}.

\subsection{IR Emission}

\begin{figure}[ht]
 \centering\leavevmode\includegraphics[width=\hsize,clip]{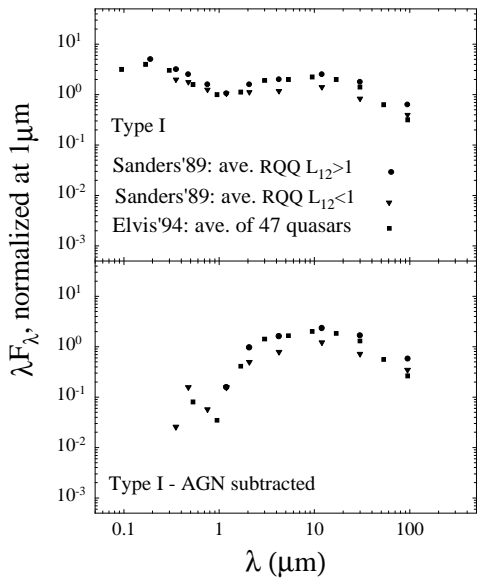}

\caption{SED's of type 1 sources. Top: Average spectra from the indicated
compilations. Bottom: The same SED's after subtracting a power law fit through
the short wavelengths ($\la$ 1\mic). These AGN-subtracted SED's are similar to
those observed in type 2 sources. \label{Fig:data1}}
\end{figure}

An obscuring dusty torus should reradiate in the IR the fraction of nuclear
luminosity it absorbs, and most AGNs do show significant IR emission. The top
panel of figure \ref{Fig:data1} shows the composite type 1 spectra from a
number of compilations. The optical/UV region shows the power law behavior
expected from a hot disk emission. At $\lambda$ \ga\ 1\mic, the SED shows the
bump expected from dust emission. The bottom panel shows the same data after
subtracting a power law fit through the short wavelengths in a crude attempt to
remove the direct AGN component and mimic the SED from an equatorial viewing of
these sources according to the unification scheme. Indeed, the AGN-subtracted
SED's resemble the observations of type 2 sources. Silicates reveal their
presence in the dust through the 10 \mic\ feature. The feature appears
generally in emission in type 1 sources and in absorption in type 2 sources
\citep{Hao07}.

The torus dust emission has been resolved recently in 8--13 \mic\
interferometry with the VLTI. The first torus was detected in NGC 1068 by
\cite{Jaffe04}, the second in Circinus by \cite{Tristram07}. The latter
observations show evidence for the long anticipated clumpy structure. The dust
temperature distributions deduced from these observations indicate close
proximity between hot ($>$ 800 K) and much cooler (\about\ 200--300 K) dust.
Such behavior is puzzling in the context of smooth-density calculations but is
a natural consequence of clumpy models \citep{AGN1}.

\subsection{Torus Size}
\label{sec:size}

Obscuration statistics provide an estimate of the torus angular width $\sigma =
\tan^{-1}H/\Ro$ (fig.\ \ref{Fig:Smooth_Clumpy}); the obscuration does not
depend individually on either $H$ or \Ro, only on their ratio. To determine an
actual size one must rely on the torus emission. The torus inner radius, \Rd,
is set by dust sublimation as
\eq{\label{eq:Rd}
    \Rd\ \simeq 0.4\,L_{45} \ \hbox{pc},
}
where $L_{45}$ is the bolometric luminosity in units of \E{45}\,\erg, and
high-resolution IR observations trace the torus to an outer radius \Ro\ \about\
5--10\,\Rd; there is no compelling evidence that torus clouds beyond \about\
20--30\,\Rd\ need be considered \citep[for details, see][]{AGN2}. A
conservative upper bound on the torus outer radius is then \Ro\ \la\
12\,$L_{45}^{1/2}$ pc.

\subsection{Torus Orientation and the Host Galaxy}
\label{sec:orientation}

Since the active region is a galactic nucleus, some obscuration can arise from
the host galaxy. However, although galactic obscuration affects some individual
sources, the strong orientation-dependent absorption cannot be generally
attributed to the host galaxy because the AGN axis, as traced by the jet
position angle, is randomly oriented with respect to the galactic disk in
Seyfert galaxies \citep{Kinney00} and the nuclear dust disk in radio galaxies
\citep{Schmitt02}. In addition, \cite{Guainazzi01} find that heavily obscured
AGN reside in a galactic environment that is as likely to be `dust-poor' as
`dust-rich'.

\section{NGC 1068---Case Study of the Molecular Environment}

NGC 1068 is the archetype Seyfert 2 galaxy and one of the most studied active
nuclei. While the galaxy is oriented roughly face-on, its AGN torus is edge-on
to within \about\,5\deg, as indicated by the geometry and kinematics of both
water maser \citep{Greenhill97, Gallimore01} and narrow-line emission
\citep{Crenshaw00}. The masers trace the disk to \la\ 1pc, inside the torus
which is traced by IR to \about\ 2--3 pc \citep{Jaffe04}. A nearly edge-on
molecular structure is traced to distances much larger than the torus outer
edge. \cite{Galliano03} model the CO and H$_2$ emission from the central region
with a clumpy molecular disk tilted roughly 15\deg\ from edge on. The disk has
a radius of 140 pc and scale height of 20 pc, for $H/R$ \about\ 0.15. From CO
velocity dispersions, \cite{Schinnerer00} find the same $H/R$ \about\ 0.15 at
$R \simeq$ 70 pc. Thus, although resembling the putative torus, the detected
molecular clouds are located in a thin disk-like structure, that does not meet
the unification scheme requirement $H/R$ \about\ 1, outside the torus. The
molecular disk orientation is much closer to the AGN than the host galaxy.
Recent adaptive optics observations by \cite{Mueller08} show evidence for
molecular clumps in infall centered on the nucleus, on a scale of only \about\
10\,pc. Imaging polarimetry at 10\mic\ by \cite{Packham07} shed some light on
the continuity between the torus and the host galaxy's nuclear environment.

These observations suggest that the AGN in NGC 1068 might be at the center of a
disk that extends beyond 100 pc, misaligned with the galactic disk. While this
central disk has a warped structure, it maintains its orientation to within
\about\,15\deg\ and thickness ($H/R$) to within \about\,0.15. The disk outer
regions are traced by molecular emission, the very inner regions (\la\ \E3\
Schwarzchild radii) provide the direct accretion channel for the black hole. In
between, the broad lines region (BLR) and torus clouds form structures much
thicker than the disk.

\section{The BLR/TOR Continuity}

Two different types of observations show that the torus is a smooth
continuation of the broad lines region, not a separate entity. IR reverberation
observations by \cite{Suganuma06} measure the time lag of the dust radiative
response to temporal variations of the AGN luminosity, determining the torus
innermost radius. Their results show that this radius scales with luminosity as
$L^{1/2}$ and is uncorrelated with the black hole mass, demonstrating that the
torus inner boundary is controlled by dust sublimation (eq.\ \ref{eq:Rd}), not
by dynamical processes. Moreover, in each AGN for which both data exist, the IR
time lag is the upper bound on all time lags measured in the broad lines, a
relation verified over a range of \E6\ in luminosity. This finding shows that
the BLR extends outward all the way to the inner boundary of the dusty torus,
validating the \cite{Netzer_Laor} proposal that the BLR size is bounded by dust
sublimation. The other evidence is the finding by \cite{Risaliti02} that the
X-ray absorbing columns in Seyfert 2 galaxies display time variations caused by
cloud transit across the line of sight. Most variations come from clouds that
are dust free because of their proximity ($<$ 0.1 pc) to the AGN, but some
involve dusty clouds at a few pc. Other than the different time scales for
variability, there is no discernible difference between the dust-free and dusty
X-ray absorbing clouds, nor are there any gaps in the distribution. These
observations show that the X-ray absorption, broad line emission and dust
obscuration and reprocessing are produced by a single, continuous distribution
of clouds. The different radiative signatures merely reflect the change in
cloud composition across the dust sublimation radius \Rd. The inner clouds are
dust free. Their gas is directly exposed to the AGN ionizing continuum,
therefore it is atomic and ionized, producing the broad emission lines. The
outer clouds are dusty, therefore their gas is shielded from the ionizing
radiation, and the atomic line emission is quenched. Instead, these clouds are
molecular and dusty, obscuring the optical/UV emission from the inner regions
and emitting IR. Thus the BLR occupies $r < \Rd$ while the torus is simply the
$r > \Rd$ region. Both regions absorb X-rays, but because most of the clouds
along each radial ray reside in its BLR segment, that is where the bulk of the
X-ray obscuration is produced. Since the unification torus is just the outer
portion of the cloud distribution and not an independent structure, it is
appropriate to rename it the TOR for Toroidal Obscuration Region.

\subsection{Outflow origin for the BLR/TOR?}

The merger of the BLR ionized clouds and TOR dusty clouds into a single
population fits naturally into a scenario first proposed by \cite{Emmering92},
involving the outflow of clouds embedded in a hydromagnetic disk wind. The
clouds are accelerated by the system rotation along magnetic field lines
anchored in the disk, as first described by \cite{Blandford_Payne}. In this
approach the TOR is that region in the wind which happens to provide the
required toroidal obscuration because the clouds there are dusty and optically
thick. The mounting evidence for cloud outflow with the geometry and kinematics
of disk winds \citep[e.g.,][]{Elvis_winds, Gallagher07} is in accord with the
outflow paradigm.

The wind scenario for the AGN toroidal obscuration unifies naturally the
different components of the system \citep{Elitzur_Shlosman}. The AGN accretion
disk appears to be fed by a midplane influx of cold, clumpy material
\citep[][and references therein]{Shlosman90}. Approaching the center,
conditions for developing hydromagnetically- or radiatively-driven winds above
this equatorial inflow become more favorable. The disk-wind rotating geometry
provides a natural channel for angular momentum outflow from the disk
\citep{Blandford_Payne} and is found on many spatial scales, from protostars to
AGN. The composition along each streamline reflects the origin of the outflow
material at the disk surface. The disk outer regions are dusty and molecular,
as observed in water masers in some edge-on cases \citep{Greenhill05}, and
clouds uplifted from these outer regions feed the TOR. Such clouds have been
detected in water maser observations of Circinus and NGC 3079. The Circinus
Seyfert 2 core provides the best glimpse of the AGN dusty/molecular component.
Water masers trace both a Keplerian disk and a disk outflow
\citep{Greenhill03}. Dust emission at 8--13\mic\ shows a disk embedded in a
slightly cooler and larger, geometrically thick torus \citep{Tristram07}. The
dusty disk coincides with the maser disk in both orientation and size. The
outflow masers trace only parts of the torus. The lack of full coverage can be
attributed to the selectivity of maser operation---strong emission requires
both pump action to invert the maser molecules in individual clouds and
coincidence along the line of sight in both position and velocity of two maser
clouds \citep{Kartje99}. In NGC~3079, four maser features were found
significantly out of the plane of the maser-traced disk yet their line-of-sight
velocities reflect the velocity of the most proximate side of the disk.
\cite{Kondratko05} note that this can be explained if, as proposed by
\cite{Kartje99}, maser clouds rise to high latitudes above the rotating
structure while still maintaining, to some degree, the rotational velocity
imprinted by the parent disk. Because the detected maser emission involves
cloud-cloud amplification that requires precise alignment in both position and
velocity along the line-of-sight, the discovery of four high-latitude maser
features implies the existence of many more such clouds partaking in the
outflow in this source.  Moving  inward from the dusty/molecular regions, at
some smaller radius the dust is destroyed and the disk composition switches to
atomic and ionized, producing a double-peak signature in some emission line
profiles \citep{Eracleous04}. The outflow from the atomic/ionized inner region
feeds the BLR and produces many atomic line signatures, including evidence for
the disk wind geometry \citep{Hall03}.

The outflow paradigm requires only the accretion disk and the clumpy outflow it
generates.  Cloud radial distance from the AGN center and vertical height above
the accretion disk explain the rich variety of observed radiative phenomena. In
both the inner and outer outflow regions, as the clouds rise and move away from
the disk they expand and lose their column density, limiting the vertical scope
of X-ray absorption, broad line emission and dust obscuration and emission. The
result is a toroidal geometry for both the BLR and the TOR. With further rise
and expansion, the density decreases so the ionization parameter increases,
turning the clouds into members of the warm absorber population. Similar
considerations can explain the broad absorption lines observed in some quasars
\citep[BAL/QSO;][]{Gallagher07}.

\subsection{TOR Cloud Properties}
\label{sec:TOR clouds}

From IR modeling, the optical depth of individual clouds should lie in the
range \tV\ \about\ 30--100 \citep{AGN2}. Assuming standard dust-to-gas ratio,
the cloud column density is \NH\ \about\ \E{22}--\E{23} \cs. The cloud density
can be constrained separately because it must be able to withstand the
black-hole tidal shearing effect. Resistance to tidal shearing sets a lower
limit on the density of the cloud \nc, leading to upper limits on its size \Rc\
and mass \Mc\ as follows \citep{Elitzur_Shlosman}:
\begin{eqnarray}
\label{eq:cloud}
   \nc &\ga& \E7{\MBH\over r_{\rm pc}^3} \ \cc,               \nonumber \\
   \Rc &\la& \E{16}{N_{\rm H,23}r_{\rm pc}^3 \over \MBH}\, {\rm cm},  \\
   \Mc &\la& 7\x\E{-3}N_{\rm H,23} R_{16}^2\,\Mo           \nonumber
\end{eqnarray}
In these expressions $r_{\rm pc}$ is the distance in pc from a black-hole with
mass \MBH\ in \E7\,\Mo, $n_7 = \nc/(\E7\,\cc)$, $N_{\rm H,23} =
\NH/(\E{23}\,\cs)$ and $R_{16} = \Rc/(\E{16}\,\rm cm)$. The resistance to tidal
shearing does not guarantee confinement against the dispersive force of
internal pressure. Self-gravity cannot confine these clouds because their low
masses provide gravitational confinement only against internal motions with
velocities \la\ 0.1 \kms. A corollary is that these clouds cannot collapse
gravitationally to form stars. While self-gravity cannot hold a cloud together
against dispersal, an external magnetic field $B \sim 1.5\,\sigma_5n_7^{1/2}$
mG would suffice if the internal velocity dispersion is $\sigma_5\x$1\,\kms.
Clouds with these very same properties can explain the masers detected in
Circinus and NGC~3079, adding support to the suggestion that the outflow water
masers are yet another manifestation of the dusty, molecular clouds that make
up the torus region of the disk-wind. Proper motion measurements and
comparisons of the disk and outflow masers offer a most promising means to
probe the structure and motion of TOR clouds.

The Resistance to tidal shearing dictates that the density inside the cloud
increase as $1/r^3$ as it gets closer to the central black-hole (eq.\
\ref{eq:cloud}). A cloud of a given density can exist only beyond a certain
radial distance; only denser clouds can survive at smaller radii, leading to a
radially stratified structure. Similarly, the Keplerian velocity increases as
$1/r^{1/2}$ as the black-hole is approached. Typical BLR cloud densities and
velocities occur at $r$ \about\ \E{16}--\E{17} cm. It is possible that the BLR
inner boundary occurs where the clouds can no longer overcome the black-hole
tidal shearing.

\subsection{The AGN Low-Luminosity End}
\label{sec:lowL}

The total TOR mass outflow rate, calculated using the properties of single
clouds listed in eq.\ \ref{eq:cloud}, exceeds its accretion rate when $L\ \la$
\E{42} \erg. Therefore, a key prediction of the wind scenario is that the torus
disappears at low bolometric luminosities because mass accretion can no longer
sustain the required cloud outflow rate, i.e., the large column densities
\citep{Elitzur_Shlosman}. Observations seem to corroborate this prediction. In
an HST study of a complete sample of low-luminosity ($\la$ \E{42} \erg) FR I
radio galaxies, \cite{Chiaberge99} detected the compact core in 85\% of
sources. Since the radio selection is unbiased with respect to the AGN
orientation, FR I sources should contain similar numbers of type 1 and type 2
objects, and Chiaberge et al suggested that the high detection rate of the
central compact core implies the absence of an obscuring torus. This suggestion
was corroborated by \cite{Whys04} who demonstrated the absence of a dusty torus
in M87, one of the sources in the FR I sample, by placing stringent limits on
its thermal IR emission. Observations by \cite{Perlman07} further solidified
this conclusion. The behavior displayed by M87 appears to be common in FR I
sources. Van der Wolk et al (private communication) performed high resolution
12 \mic\ imaging observations of the nuclei of 27 radio galaxies with the VISIR
instrument on the VLT. These observations provide strong confirmation of the
torus disappearance in FR I sources.  They show that all the FR I objects in
the sample lack dusty torus thermal emission, although they have non-thermal
nuclei.  Thermal dust emission was detected in about half the FR II nuclei,
which generally have higher luminosities. In contrast, in almost all broad line
radio galaxies in the sample the observations detected the thermal nucleus.
Significantly, \cite{Ogle07} find that most FR I and half of FR II sources have
$L/\LEdd < 3\cdot\E{-3}$, while all sources with broad Balmer lines have
$L/\LEdd > 3\cdot\E{-3}$ (note that \LEdd\ = \E{45}\MBH\, \erg).

The AGNs known as LINERs provide additional evidence for the torus
disappearance. \cite{Maoz05} conducted UV monitoring of LINERs with $L$ \la\
\E{42} \erg\ and detected variability in most of them.  This demonstrates that
the AGN makes a significant contribution to the UV radiation in each of the
monitored sources and that it is relatively unobscured in all the observed
LINERs, which included both type 1 and type 2. Furthermore, the histograms of
UV colors of the type 1 and 2 LINERs show an overlap between the two
populations, with the difference between the histogram peaks corresponding to
dust obscuration in the type 2 objects of only \about\ 1 magnitude in the R
band. Such toroidal obscuration is minute in comparison with the torus
obscuration in higher luminosity AGN. The predicted torus disappearance at low
$L$ does not imply that the cloud component of the disk wind is abruptly
extinguished, only that its outflow rate is less than required by the IR
emission observed in quasars and high-luminosity Seyferts. When the outflow
drops below these ``standard'' torus values, the outflow still provides
toroidal obscuration as long as its column exceeds \about\ \E{21} \cs. Indeed,
Maoz et al find that some LINERs do have obscuration, but much smaller than
``standard''. Line transmission through a low-obscuration torus might also
explain the low polarizations of broad H$\alpha$ lines observed by
\cite{Barth99} in some low luminosity systems.

If the toroidal obscuration were the only component removed from the system,
all low luminosity AGN would become type 1 sources. In fact, among the LINERs
monitored and found to be variable by Maoz et al there were both sources with
broad H$\alpha$ wings (type 1) and those without (type 2). Since all objects
are relatively unobscured, the broad line component is truly missing in the
type 2 sources in this sample. Similarly, \cite{Panessa02} note that the BLR is
weak or absent in low luminosity AGN, and \cite{Laor03} presents arguments that
some ``true'' type 2 sources, i.e., having no obscured BLR, do exist among AGNs
with $L$ \la\ \E{42}\,\erg. The absence of broad lines in these sources cannot
be attributed to toroidal obscuration because their X-ray emission is largely
unobscured. These findings have a simple explanation if when $L$ decreases
further beyond the disappearance of the TOR outflow, the suppression of mass
outflow spreads radially inward from the disk's dusty, molecular region into
its atomic, ionized zone. Then the torus disappearance, i.e., removal of the
toroidal obscuration by the dusty wind, would be followed by a diminished
outflow from the inner ionized zone and disappearance of the BLR at some lower
luminosity. Indeed, a recent review by Ho (2008) presents extensive
observational evidence for the disappearance of the torus and the BLR in low
luminosity AGN.  Such an inward progression of the outflow turnoff as the
accretion rate decreases can be expected naturally in the context of disk winds
because mass outflow increases with the disk area. A diminished supply of
accreted mass may suffice to support an outflow from the inner parts of the
disk but not from the larger area of its outer regions. And with further
decrease in inflow rate, even the smaller inner area cannot sustain the disk
outflow. Since the accreted mass cannot be channeled in full into the central
black hole, the system must find another channel for release of the excess
accreted mass, and the only one remaining is the radio jets. Indeed,
\cite{Ho02} finds that the AGN radio loudness $\R = L_{\rm radio}/L_{\rm opt}$
is {\em inversely} correlated with the mass accretion rate $L/\LEdd$. This
finding is supported by \cite{Sikora07}, who have greatly expanded this
correlation and found an intriguing result: \R\ indeed increases inversely with
$L/\LEdd$, but only so long as $L/\LEdd$ remains \ga\ \E{-3}. At smaller
accretion rates, which include all FR I radio galaxies, the radio loudness
saturates and remains constant at \R\ \about \E4. This is precisely the
expected behavior if as the outflow diminishes, the jets are fed an
increasingly larger fraction of the accreted mass and finally, once the outflow
is extinguished, all the inflowing material not funneled into the black hole is
channeled into the jets, whose feeding thus saturates at a high conversion
efficiency of accreted mass. It is important to note that radio loudness
reflects the relative contribution of radio to the overall radiative emission;
a source can be radio loud even at a low level of radio emission if its overall
luminosity is small, and vice versa.

\begin{figure}[ht]
 \centering\leavevmode\includegraphics[width=\hsize,clip]{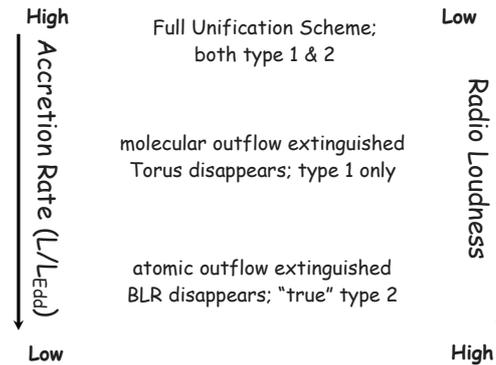}
\caption{Conjectured scheme for AGN evolution with decreasing accretion rate.}
\label{Fig:AGN_scheme}
\end{figure}

The evolutionary scheme just outlines is sketched in figure
\ref{Fig:AGN_scheme}. A similar anti-correlation between radio loudness and
accretion rate exists also in X-ray binaries. These sources display switches
between radio quiet states of high X-ray emission and radio loud states with
low X-ray emission \citep{Fender04}. While in X-ray binaries this behavior can
be followed with time in a given source, in AGN it is only manifested
statistically. Comparative studies of AGN and X-ray binaries seem to be a most
useful avenue to pursue.

\begin{acknowledgements}
Partial support by NSF and NASA is gratefully acknowledged.
\end{acknowledgements}

\end{document}